# Wafer scale growth and characterization of edge specific graphene nanoribbons

*Alexei A.Zakharov\*[⊥], Nikolay A.Vinogradov[⊥], Johannes Aprojanz[+], Christoph Tegenkamp[§+],*

*Claudia Struzzi[⊥], Tikhomir Yakimov[ǂ], Rositsa Yakimova[ǂ] and Valdas Jokubavicius[ǂ]*

[⊥]MAXIV Laboratory, Lund University, Fotongatan 2, 22484, Lund, Sweden

[+]Institut für Festkörperphysik, Leibniz Universität Hannover, Appelstr.2, D-030167 Hannover, Germany

[§]Institut für Physik, Technische Universität Chemnitz, Reichenhainer Str. 70, 09126, Chemnitz, Germany

[ǂ]IFM Linköping University, SE-58183, Linköping, Sweden

Abstract

One of the ways to use graphene in field effect transistors is to introduce a band gap by quantum confinement effect [1]. That is why narrow graphene nanoribbons (GNRs) with width less than 50nm are considered to be essential components in future graphene electronics. The growth of graphene on sidewalls of SiC(0001) mesa structures using scalable photolithography was shown to produce high quality GNR with excellent transport properties [2-7]. Such epitaxial graphene nanoribbons are very important in fundamental science but if GNR are supposed to be used in advanced nanoelectronics, high quality thin (<50nm) nanoribbons should be produced on a large (wafer) scale. Here we present a technique for scalable template growth of high quality GNR on Si-face of SiC(0001) and provide detailed structural information along with transport properties. We succeeded to grow GNR along both [1$\bar{1}$00] and [11$\bar{2}$0] crystallographic directions. The quality of the grown nanoribbons was confirmed by comprehensive characterization with high resolution STM, dark field LEEM and transport measurements.

First attempts to produce GNR by standard lithography methods [8-10] turned out to severely suffer from the pattering process resulting in disordered and rough edges. These defects significantly control the electronic properties of GNR, e.g. the charge carriers have very small electron mean free path resulting in quite high (~1kΩ/□) sheet resistance [11-13]. That is why the focus was shifted to a direct growth of nanoribbons which avoids deleterious post-processing. Most promising way for producing graphene nanostructures is selective growth of graphene by sublimation on the side walls of SiC(0001) mesa structures. On thermal etching/annealing the vertical side wall reconstructs into facets angled at 20-30 deg. relative to the surface plane. Due to the different bonding of the Si atoms, graphene forms predominantly on the facets. Therefore, it is possible to grow isolated graphene nanoribbons while keeping much less conducting buffer layer on top and bottom terraces. 1D ballistic transport was observed at room temperature with electron mean free path length reaching tens of micrometers [5]. The direction of the graphene ribbon edge can be controlled by using mesa's trenches in SiC oriented either parallel or perpendicular to the SiC [1$\bar{1}$00] direction. If the trench is oriented parallel to the [1$\bar{1}$00] direction, graphene nanoribbons have their zig-zag (ZZ) edge along the trench and such ribbons will be referred as ZZ nanoribbons. Conversely, in the perpendicular ([11$\bar{2}$0]) direction, GNR has its armchair (AC) edge along the trench and such ribbons are called as AC nanoribbons. The properties of ZZ and AC nanoribbons are markedly different due to their different electronic band structure and edges and for the first time we provide a detailed description for both orientations grown under identical conditions on the same substrate.

Semi-insulating and N-type doped 6H-SiC(0001) substrates were used as template for graphene growth. Mesa structures (20-30nm height) were defined using a combination of standard UV lithography and RIE etching (P=30W, 20/7 $SF_6/O_2$, 0.05 mbar) on 8x8 $mm^2$ samples (see Fig.1a). Before graphene growth, an extra step (thermal etching) to transfer the mesa sidewalls into inclined facets was included. This step was also necessary for the doped wafers to remove polishing scratches. The semi-insulating substrates have epi-ready Si –face and they were used with and without thermal etching since faceting also occur *in situ* during graphene growth. Thermal etching of mesa structures fabricated on SiC surfaces was done in a graphite crucible which is inductively heated. Thermal etching temperatures vary in a range of 1700 - 1800$^o$C. Inside the graphite crucible a Ta foil and a 1 mm thick graphite spacer with an opening and the SiC sample with the surface to be etched facing the Ta foil are placed on top of each other (see Fig.S1, Supporting info and ref[14]). At elevated temperatures Ta reacts with carbon bearing species which are sublimed from the SiC and forms stable carbide [15]. Upon such reaction the ambient between the SiC and Ta is enriched with silicon and graphitization of SiC surface is prevented. A medium vacuum (E-5mbar) is maintained during the etching process to enhance the etching rate. In our previous study we demonstrated that the etching rate of 4H-SiC (0001) surface at 1800$^o$C in vacuum using the same arrangement is about 190 nm/h[14]. Therefore, to transfer the mesa sidewalls into inclined facets or to remove residual polishing scratches the etching time of 1 min is usually used.

For the growth of graphene nanoribbons, the samples were annealed in an inductively heated furnace at 1800°C with a ramping rate of 25°C/min under 850 mbar Argon atmosphere for 1min and immediately cooled down after that [16]. These conditions are conducive to fast surface kinetics due to the high temperature while also favoring a low rate of silicon loss from the surface, leading to larger areas of homogeneous graphene than high vacuum and ultra-high vacuum (UHV) growth. Very strict control of the growth temperature and duration is required in order to ensure local formation of nanoribbons on side walls while keeping the terraces free of graphene patches. LEEM and micro-LEED studies were performed at aberration corrected low energy electron microscope (Elmitec GmbH) installed at MAXPEEM beamline (MAXIV synchrotron, Lund, Sweden). Scanning tunneling microscopy (STM) data were collected at a commercial VT-STM from Omicron, GmbH at room temperature. Prior to performing LEEM and STM studies, the samples were annealed in the ultra-high vacuum (base pressure in both systems 5-7E-11mbar) to $600^{o}$C for 15 minutes. Electrochemically etched W tip (wire diameter = 0.38 mm) was used for STM imaging. In order to comply with abrupt height variations across the sample (mesa facets) the sample was scanned along the facets. That is, the tip was scanning a line along the facet where the height variations were small, and then moving to the next line at a different height. In such a way, atomically-resolved images of GNRs on a steep wall were obtained.

In situ transport measurements were performed in a commercial Omicron Nanoprobe-STM/SEM system, which also hosts a SPA-LEED. All transport measurements were carried out at room temperature using chemically etched Tungsten tips.

Fig. 1b depicts large area SEM images taken at three different areas which were 2 mm apart from each other. The mesa structures and GNR (yellow lines) are clearly visible. Besides some dark patches of excess monolayer graphene, the variation of the SEM contrast level is very low demonstrating a remarkable homogeneity of the growth stage on the millimeter scale.

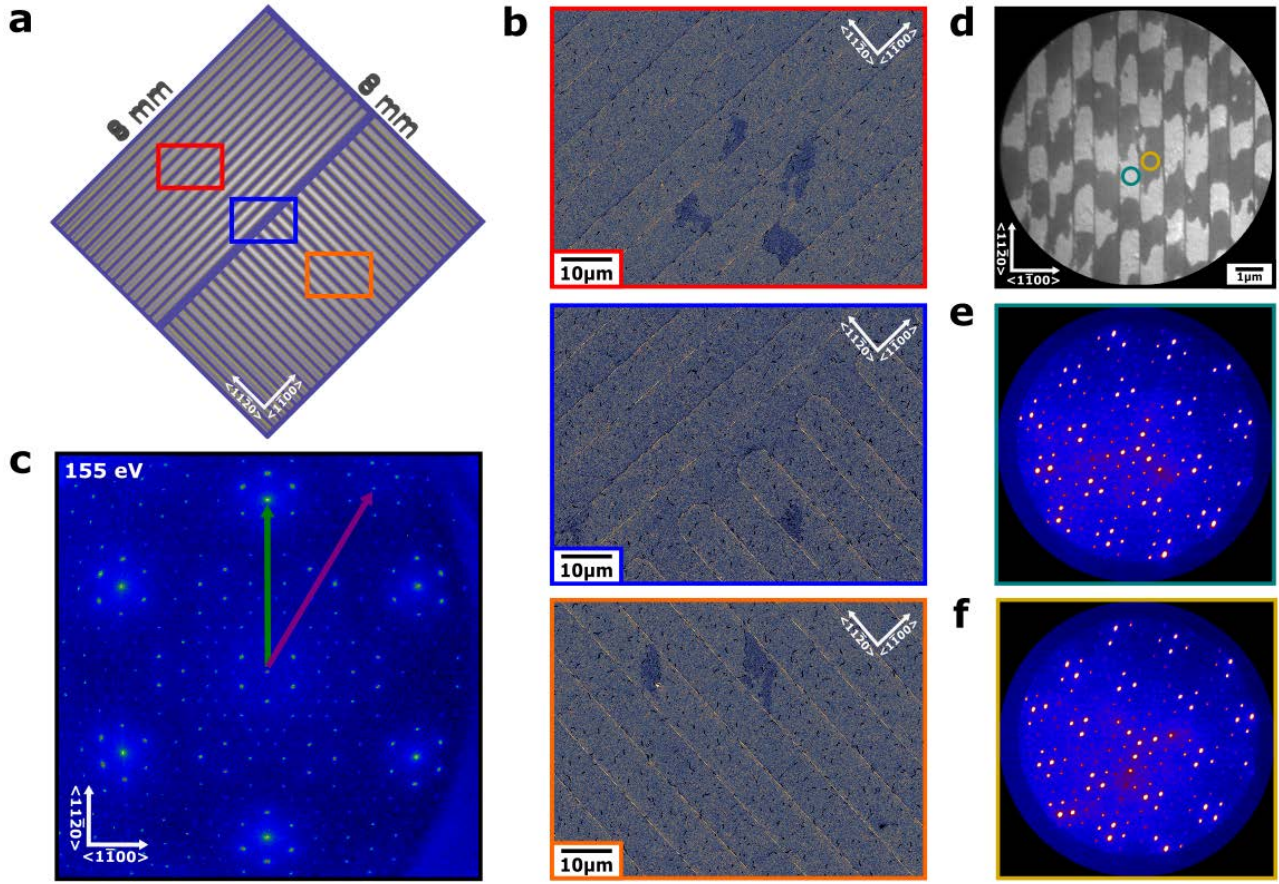

Fig.1. a) Schematic view of RIE structured samples, 8x8 mm^2 in size. The samples host mesa structures oriented along the $[11\bar{2}0]$ (armchair) and $[1\bar{1}00]$ (zigzag) direction, respectively. b) SEM images taken at positions highlighted in a) showing the mesa structures after GNR growth. Almost no variation of the SEM contrasts levels across the entire surface demonstrating a remarkable homogeneity. c) SPA-LEED pattern taken at 155 eV characteristic for the buffer layer grown on SiC. The first order spots of SiC (green) and graphene (violet) are indicated by arrows. d) Tilted bright field (TBF) (E=23eV) LEEM image showing a perfect buffer layer on the top and bottom terraces with two domains originating from underlying substrate SiC after GNR growth, FoV=10μm; e,f) μ-LEED images from two single domains in the TBF LEEM (d) marked with the circles.

SPA-LEED is an integrating method (probing area 800 µm$^2$) in order to characterize the crystallographic structure of surfaces. A representative SPA-LEED image of a surface with a rather large mesa pitch size (8 µm) after GNR growth is shown in Fig.1c. The pattern displays first order diffraction spots of SiC (green) and graphene (violet) in coexistence with the 6√3x6√3 R30$^0$ reconstruction, distinctive for a fully-developed buffer layer on SiC. Any features originating from the facets are missing due to the small density of GNRs compared to the planar faces. LEEM and µ-LEED measurements further confirm the existence of the interfacial buffer layer on the trenches and plateaus of the mesa. The epitaxial buffer layer is of exceptional quality as can be seen from Fig. 1d where a tilted bright field LEEM image reveals two different domains inherent in SiC structure [17] and translated into the epitaxial buffer layer. The domain width is on the micrometer scale and it was possible to illuminate the entire uniform domain area to collect the diffraction pattern from a single domain. The sharp diffraction patterns at 38.7eV are shown in Fig.1e and Fig.1f demonstrating three-fold rotation symmetry which suggests single domain diffraction. So both top and bottom trenches of the mesa host a high quality buffer layer. To quantify topologically the side wall graphene a dark field LEEM mode was used. Fig.2 shows a sequence of experimental steps for dark field LEEM imaging. First, a bright field LEEM image (Fig.2a) is acquired. Then, the incoming electron beam is limited to a very small spot (in the present case 250nm) on the surface covering one of the walls. With this setup it is possible to acquire micro-LEED pattern from a very small area which includes a single wall. The intensity of such a small beam is usually very strong in the LEED mode and it leaves a mark (beam effect, Fig.2a, yellow circle) on the buffer layer after acquiring diffraction pattern. This mark serves as a double check that the LEED has been acquired from the right place (in this case on the mesa wall) to make sure that there is no drift during the course of the measurement. Corresponding μ-LEED pattern is presented in Fig.2b. The μ-LEED shows a typical buffer layer pattern with $6\sqrt{3} \times 6\sqrt{3}$ R$30^0$ symmetry. In addition to the $6\sqrt{3}$ spots of the buffer layer, there are spots that move with energy (insets). These spots are associated with electron diffraction from the faceted walls. The trenches in Fig.2a are oriented parallel to the $[1\bar{1}00]$ direction (ZZ direction) and the moving spots are from ZZ graphene nanoribbons. The graphene nanoribbon image from the facet diffraction spot (dark field LEEM) is depicted in Fig.2c. As expected, in accordance with the facet spot in the LEED pattern, GNRs show up only on the one side of the mesa, i.e. graphene on every second wall is highlighted. The faceting and nanoribbon growth are different in the orthogonal (AC) direction (see below) though they are treated identically with ZZ ones.

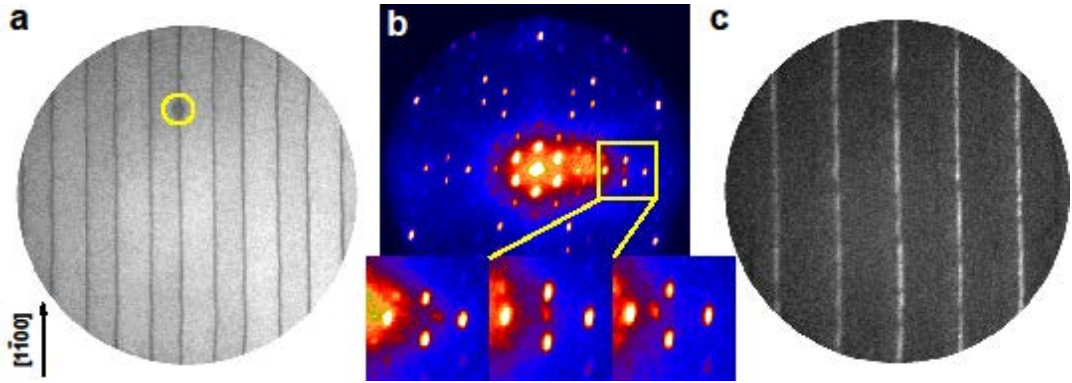

Fig.2. LEEM and micro-LEED studies of ZZ GRNs. a) Bright field LEEM image (FoV=10μm, E=5eV), yellow circle marks the sampling area (400nm) for the micro-LEED from single mesa wall; b) m-LEED pattern (E=25.5eV) from a single mesa wall. Insets in the bottom (E=25eV, 25.5eV and 26eV from left to right, correspondingly) show a moving facet spot chosen for dark-field imaging; c) LEEM DF image (E=25.5eV) highlighting graphene nanoribbons on every second wall of the mesa structure. Crystallographic orientation of the SiC substrate is in the bottom-left corner.

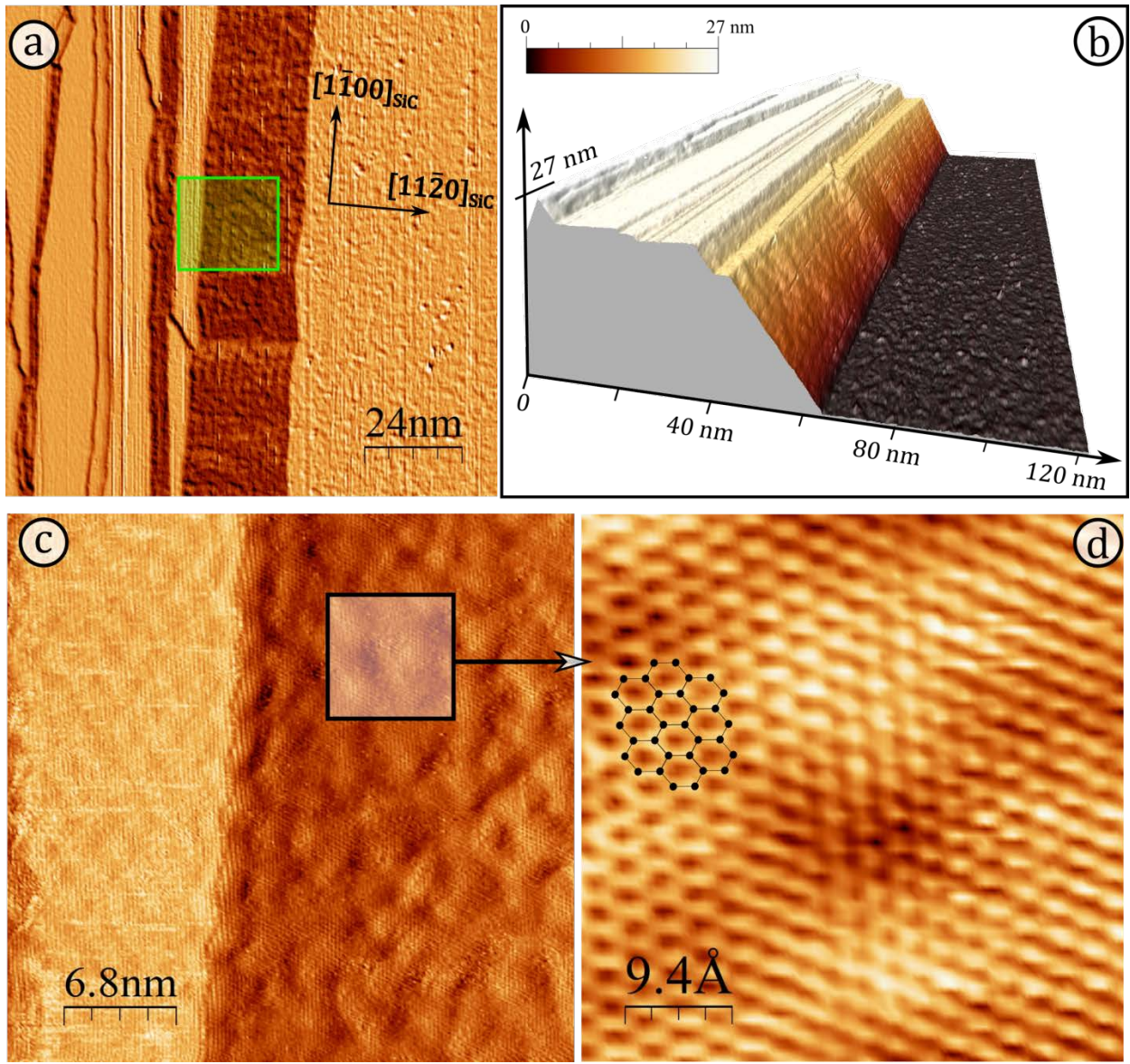

Fig.3 Topography and structure of the ZZ GNRs. *a)* a survey STM image, topography (Z-channel) differentiated. A 3D representation of the same image is shown in *b*, where the topographic information is preserved. Image size in both *a* and *b* is 120x120 nm, tunneling parameters are $V_b/I_t$ = 1.5 V/1 nA. *c)* a high-resolution STM image of the facet with a micro-step, highlighted with the green square in *a* ($I_t$-channel). Image size is 35x35 nm, $V_b/I_t$ = 0.15 V/1.75 nA. A high-resolution STM micrograph (Z-channel) of the area highlighted in *c* with black frame is presented in *d*. Image size is 48x48 Å, $V_b/I_t$ = 0.15 V/1.65 nA.

But overall, the single wall GNRs are quite homogeneous along the trench in both directions thus demonstrating that graphene ribbons self-assembled on the side walls and this growth approach is able to produced GNRs on a large scale.

To further characterize GNRs in both orientations, STM and transport measurements were performed both on ZZ and AC nanoribbons. Fig.3 shows the results of STM studies on the ZZ nanoribbons. In Fig.3a an STM overview of a typical ZZ nanoribbon together with 3D profile (Fig.3b) are presented. In order to dismiss the contrast variation from the 30-nm height jump on the facet, a differential filter was applied to the image ($I(x,y) = dZ/dx + dZ/dy$). The main graphene nanoribbon is about 30-40nm wide, resting on the mesa structure sidewall with slope between 20 and 30 degrees. It is visible by the dark contrast in the center of Fig. 3a and its 3D representation is shown in Fig. 3b. Inspecting these images further, one notices that a certain step bunching takes place in the vicinity of the top edge of the facet (the left part of image in Fig. 3a). We have observed a few (typically, one to two) mini-facets of 1-2 nm height, separated by narrow (4-20nm) and irregularly-shaped terraces. Fig. 3c shows a closer view of the main facet with a mini-terrace, highlighted in with a green square in Fig. 3a. The fast Fourier transform taken from the leftmost part of this image (SI, Fig.S3) shows a set of satellite spots characteristic for buffer layer on SiC. This observation confirms that the mini-terraces are covered with a high-quality buffer layer, and do not differ from the rest of the macro-terraces of the sample. The atomically resolved image of a ZZ GNR, demonstrated in Fig. 3d shows that indeed graphene nanoribbon is oriented with its $[1\bar{1}00]$ "zig-zag" direction along the mesa wall. Therefore, the top edge of the ribbon should maintain ZZ orientation as expected from the substrate orientation.

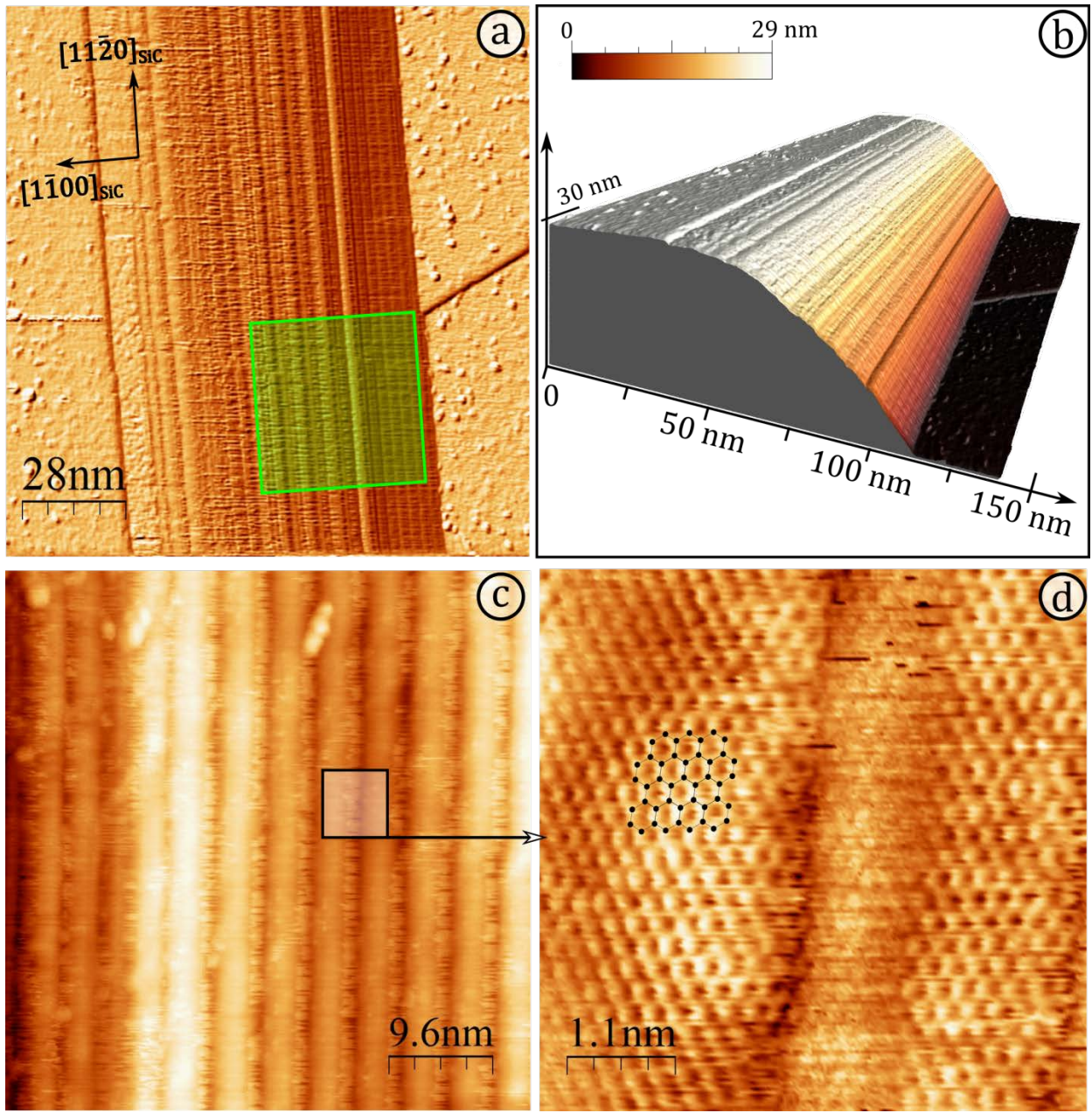

Fig.4. Topography and structure of the AC GNRs. *a)* a survey STM image, topography (Z-channel) differentiated. A 3D representation of the same image is shown in *b*, where the topographic information is preserved. Image size in both *a* and *b* is 140x140 nm, tunneling parameters are $V_b/I_t$ = 1.75 V/1.5 nA. *c)* A closer view of the facet highlighted in *a* with the green square (Z-channel). Image size is 48x48 nm, $V_b/I_t$ = 1.75 V/1.75 nA. *d* A high-resolution STM micrograph of an individual step (topography channel)

highlighted in *c* with the black square. Image size is 60x60 Å, $V_b/I_t$ = 0.15 V/1.65 nA. A schematic graphene lattice is superimposed on the image in the top-left corner.

Nanowires grown in the orthogonal (AC) direction are different in many respects compared to the ZZ ones. Fig.4 summarizes the STM results on AC GNRs. The AC mesa walls consist of a number of micro-steps, typically 10-20, of 2-3 nm width and 1-1.5 nm height. Such appearance is somewhat reminiscent of a "washboard" or a "ladder" structure, as clearly visible in Figs. 4a-c and in the µ-LEED pattern (Fig.S2b, SI) . This, in turn, has an effect on the GNR formed on such a facet. The AC GNRs appears to be strongly corrugated following the topography of the substrate, its whole width is larger (up to 100nm) with regard to the ZZ GNR and the sidewall slope is smaller (<30deg.). Nevertheless, these ribbons still demonstrate high crystallographic quality and atomically resolved STM patterns as evident from Fig. 4d. Moreover, it is important to note, that AC GNRs do not break on the micro-step edges, but rather exhibit a strong corrugation due to a remarkable flexibility and strength of graphene. A similar morphology was observed for AC nanoribbons in form of miniterraces and minifacets due to fracturing of the initial facets during the course of graphene growth [6]. Obviously, the AC facet has a higher surface energy compared to the ZZ facet which remains stable during graphene growth.

Fig. 5a shows a SPA-LEED image acquired on an area densely packed with mesas parallel to the AC direction. Again the typical buffer layer diffraction pattern is observed along with an additional stripe structure. High resolution scans in Fig. 5b taken at different energies reveal moving spots associated with the AC facets. The calculated facet angle of $\Theta = 21° \pm 1°$ nicely resembles the STM results (see Fig.5c). These findings strongly support that the debunching of the sidewall along [11$\bar{2}$0] into the washboard is a hallmark of AC facets after GNR growth and moreover the remarkable crystalline quality of the as-grown GNR.

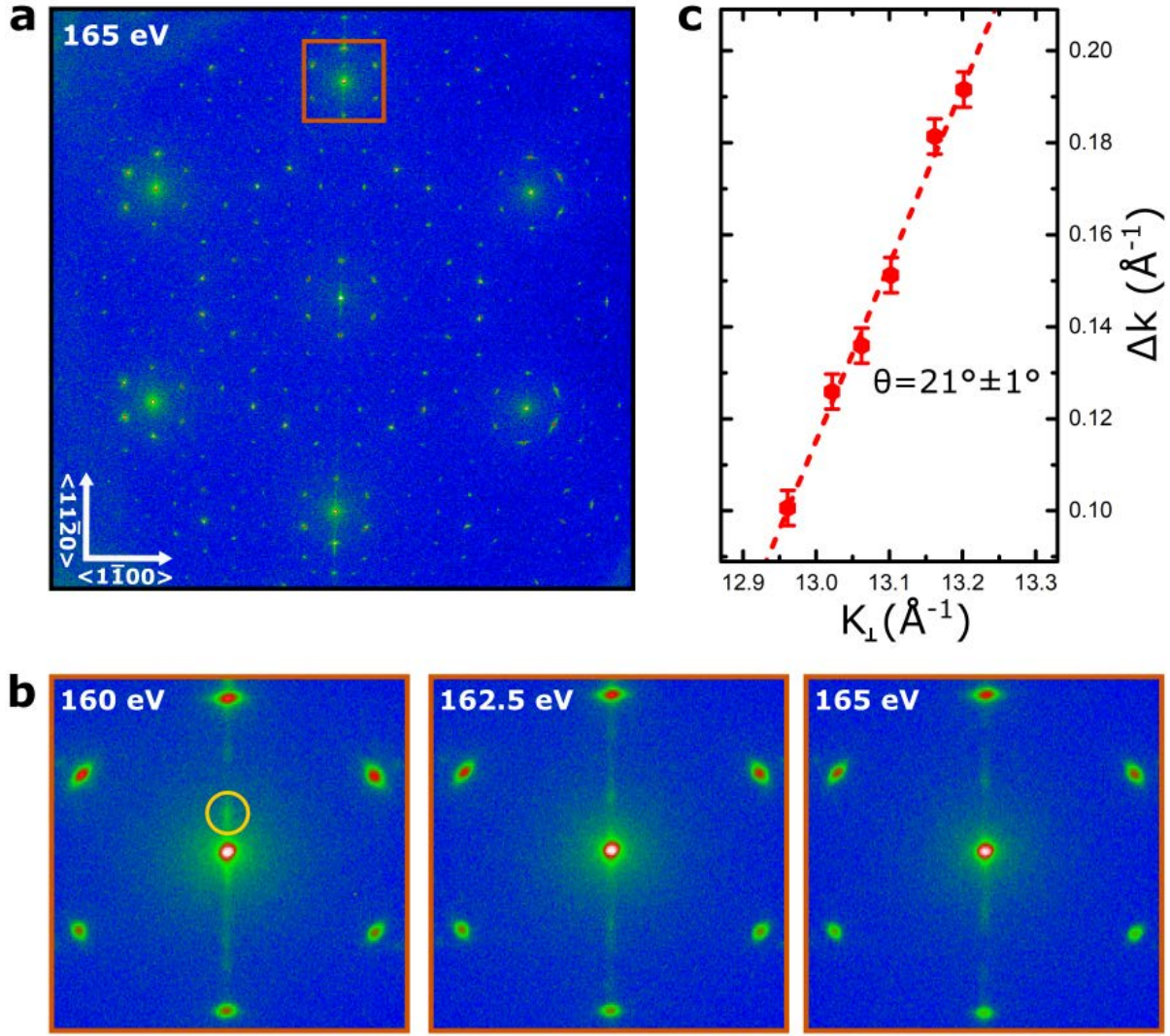

Fig.5 a) SPA-LEED image taken on an area with a high density of armchair mesa structures at 166 eV showing a similar pattern as in Fig.1 c) with additional stripes along the <11-20> direction. b) High resolution scan of marked are in a) at different energies. The additional facet spots are moving with energy. c) Extracted Δk and $K_\perp$ values. From the linear increase a facet angle of Θ = 21° ± 1° was calculated.

The local resistance of AC and ZZ GNR was measured using a linear four-point probe arrangement. The current was injected by two outer probes while the voltage was simultaneously recorded by a pair of inner probes (see inset Fig.6) ruling out any contributions from lead and contact resistances. Fig. 6 depicts exemplary IV curves taken on AC (green) and ZZ GNR (blue) with equidistant probe spacing of 1µm and 500 nm, respectively. All displayed IV curves show Ohmic characteristics, thus the corresponding resistances were obtained by a linear fit.

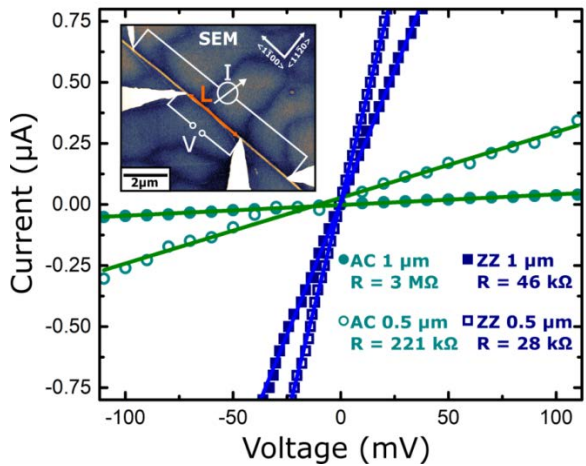

Fig.6 In-situ four-point probe transport measurements of zigzag and armchair GNR. Exemplary IV curves taken for two distinct probe spacings L on both kinds of GNR. The zigzag GNR shows a resistance close to 1 $h/e^2$ while the resistance of the armchair GNR is almost 10 times larger at the same probe spacing. The tip arrangement is depicted in the inset.

Comparing the measurements with a 1 µm probe spacing it is evident that the resistance of ZZ GNRs is significantly reduced. Alike at L = 500 nm the AC GNR display an almost 10 times higher local resistance compared to ZZ GNR. The periodic washboard structure, present on AC sidewalls, consists of an alternating sequence of epitaxial graphene on miniterraces (w ~ 3 nm) accompanied by suspended GNR at the step edges (w ~ 1 nm). Band gaps of 0.3 – 1.5 eV can exist in AC ribbons of such dimensions resulting in larger resistances compared to the metallic ZZ GNRs [18]. Interestingly, the ZZ GNRs exhibit a resistance value of $R_{ZZ,500nm}$ = 28 kΩ ~ $h/e^2$ indicating single channel ballistic transport, in full agreement with our previous findings [5,7]. In general, ballistic transport signatures were only observable when the debunching into the washboard structure was suppressed.

In summary, our results demonstrate that free standing graphene grows on the faceted mesa walls no matter what crystallographic direction of the SiC(0001) substrate is used for the patterning. The quality of the zig-zag ribbons is somewhat better compared with the arm-chair ones and these ribbons are more stable and demonstrate ballistic transport. In turn, armchair ribbons have much higher resistance which indicates possible band gap opening, property which can be used in field effect transistors. The presented pre-processing and growth technique is viable for the production of nanoribbons on a large scale and applications in graphene electronics are readily envisaged.

## ASSOCIATED CONTENT

**Supporting Information** is in a separate file (supporting info.pdf). The file contains a scketch of growth and etching reactors, LEEM and micro-LEED data on the arm-chair graphene nanoribbons and Fourier transforms of the STM atomically resolved images.

## AUTHOR INFORMATION

### Corresponding Author

*Corresponding_Author_E-mail: Alexei.Zakharov@maxiv.lu.se tel. +46(0)733439556

### Author Contributions

C.T. conceived and supervised the project, A.A.Z performed LEEM and micro-LEED measurements and wrote the manuscript. N.V. performed STM measurements, J.A. performed 4-probe STM, SEM and SPA-LEED mesurements, C.C performed LEEM mesurements, R.Y. and T.Y. grew graphene nanoribbons, V.J. performed thermal etching. All authors have discussed the results and commented on the paper.

## ACKNOWLEDGMENT

Financial support by the Deutsche Forschungsgemeinschaft (Te386/12-1 and Te 386/13-1) is gratefully acknowledged by J.A. and C.T.  A.A.Z, C.S and N.V. acknowledge Vetenskapsrådet (TAILSPIN project). R.Y. would like to acknowledge financial support by the Swedish Agency for Strategic research (SSF) via project GMT14-0077.

ABBREVIATIONS

LEEM – Low Energy Electron Microscopy, BFLEEM – Bright Field Low Energy Electron Microscopy, DFLEEM – Dark Field Low Energy Microscopy, LEED – Low Energy Electron Diffraction, SPA LEED – Spot Profile Analysis Low Energy Electron Diffraction, STM – Scanning Tunneling Microscopy, GNR – graphene Nano Ribbons

TABLE OF CONTENT GRAPHIC

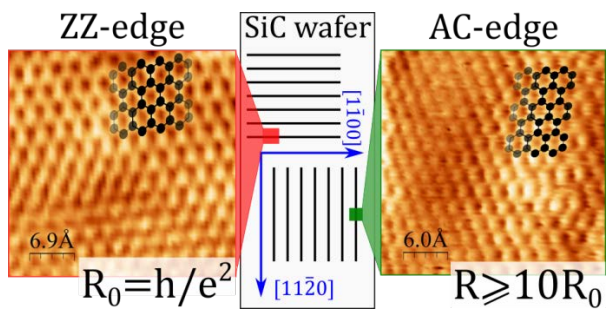